\documentclass[aip,
 amsmath,amssymb,
 preprint,%
a4paper
]{revtex4-2}

\usepackage{graphicx}
\usepackage{dcolumn}
\usepackage{bm}

\usepackage[utf8]{inputenc}
\usepackage[T1]{fontenc}
\usepackage{mathptmx}
\usepackage{etoolbox}
\usepackage[table]{xcolor}
\usepackage[colorlinks=true,linkcolor=blue,citecolor=blue,urlcolor=blue,allbordercolors=white]{hyperref}

\begin{document}
\title{\large Biological computation through recurrence}

\author{Mar\'ia Sol Vidal-Saez}
\affiliation{Department of Medicine and Life Sciences, Universitat Pompeu Fabra, Dr Aiguader 88, 08003 Barcelona, Spain}

\author{Oscar Vilarroya}
\affiliation{Department of Psychiatry and Legal Medicine, Universitat Aut\`onoma de Barcelona, 08193, Cerdanyola del Vallès, Spain}
\affiliation{Hospital del Mar Medical Research Institute (IMIM), Dr Aiguader 88, 08003, Barcelona, Spain}

\author{Jordi Garcia-Ojalvo}
\email{jordi.g.ojalvo@upf.edu}
\affiliation{Department of Medicine and Life Sciences, Universitat Pompeu Fabra, Dr Aiguader 88, 08003 Barcelona, Spain}

\begin{abstract}
One of the defining features of living systems is their adaptability to changing environmental conditions.
This requires organisms to extract temporal and spatial features of their environment, and use that information to compute the appropriate response.
In the last two decades, a growing body or work, mainly coming from the machine learning and computational neuroscience fields, has shown that such complex information processing can be performed by recurrent networks. 
Temporal computations arise in these networks through the interplay between the external stimuli and the network's internal state.
In this article we review our current understanding of how recurrent networks can be used by biological systems, from cells to brains, for complex information processing.
Rather than focusing on sophisticated, artificial recurrent architectures such as long short-term memory (LSTM) networks, here we concentrate on simpler network structures and learning algorithms that can be expected to have been found by evolution.
We also review studies showing evidence of naturally occurring recurrent networks in living organisms.
Lastly, we discuss some relevant evolutionary aspects concerning the emergence of this natural computation paradigm.
\end{abstract}

\maketitle
\section{Introduction}

The survival of any living system depends on its capacity to sense multiple signals in its surroundings, integrate those external signals, and activate an adequate response.
Not only the spatial, but also the temporal structure of these signals provide key information to the organisms \cite{Buonomano1995}. 
Temporal information, in particular, is crucial to anticipate changes in the environment.
In these cases, having some memory mechanism is clearly beneficial.
In the mammalian brain, for instance, processing the temporal structure of visual stimuli gives information on the direction and speed of objects, as well as on the duration of and time between events \cite{Buonomano2009}.
In the somatosensory system, the contribution of temporal encoding is not limited to dynamic aspects of the stimulus, such as motion detection, but is also involved in stationary aspects such as object and texture discrimination \cite{Ahissar2001}.
It has been postulated that recurrence due to extensive feedback can underlie the temporal information-processing abilities of cortical neural networks \cite{Destexhe2006,Buonomano2009}. 

The importance of processing temporal information is not limited to neural systems. 
Non-neural individual cells such as bacteria possess the capability to associate groups of events that commonly co-occur or follow a particular sequence, thus exhibiting a form of associative memory \cite{balaban13}.
As an illustration, \textit{Escherichia coli} anticipates itself for low oxygen environments upon detecting an increase in temperature, which would signal ingestion by a mammal \cite{Tagkopoulos2008}.
Similarly, \textit{Candida albicans} cells, a fungal pathogen, upregulates oxidative stress resistance genes upon detection of an increase in glucose levels (which indicates they have entered a host), preemptively reacting to its immune system \cite{Schild2007}. 

Apart from this associative memory capacity, recent studies indicate that microorganisms display both short- and long-term memory. For instance, the stress response of \textit{Bacillus subtilis} relies not solely on its present growth conditions but also on its past \cite{Wolf2008}. The integration and storage of previous history, \textit{i.e.} memory, in bacterial cells is a subject of considerable interest.
It has been suggested that this capability of bacteria is dependent on a complex network of transcriptional interactions  \cite{Lee2002,MartnezAntonio2003}. 
These types of networks are called gene regulatory networks.

Gene regulatory networks from multiple organisms have been found to possess a structured recurrent architecture reminiscent of the one underlying a computational paradigm known as reservoir computing (RC) \cite{GabaldaSagarra2018}. In this scheme, temporal computations emerge from the inherent structure and dynamics of the network a very efficient learning process \cite{Jaeger2001original,Verstraeten2007}. This paradigm was also proposed to underlie brain computations under the name of liquid-state machines (LSMs) \cite{Maassoriginal}.

\begin{figure}[htbp]
\resizebox{0.5\textwidth}{!}{%
 \includegraphics{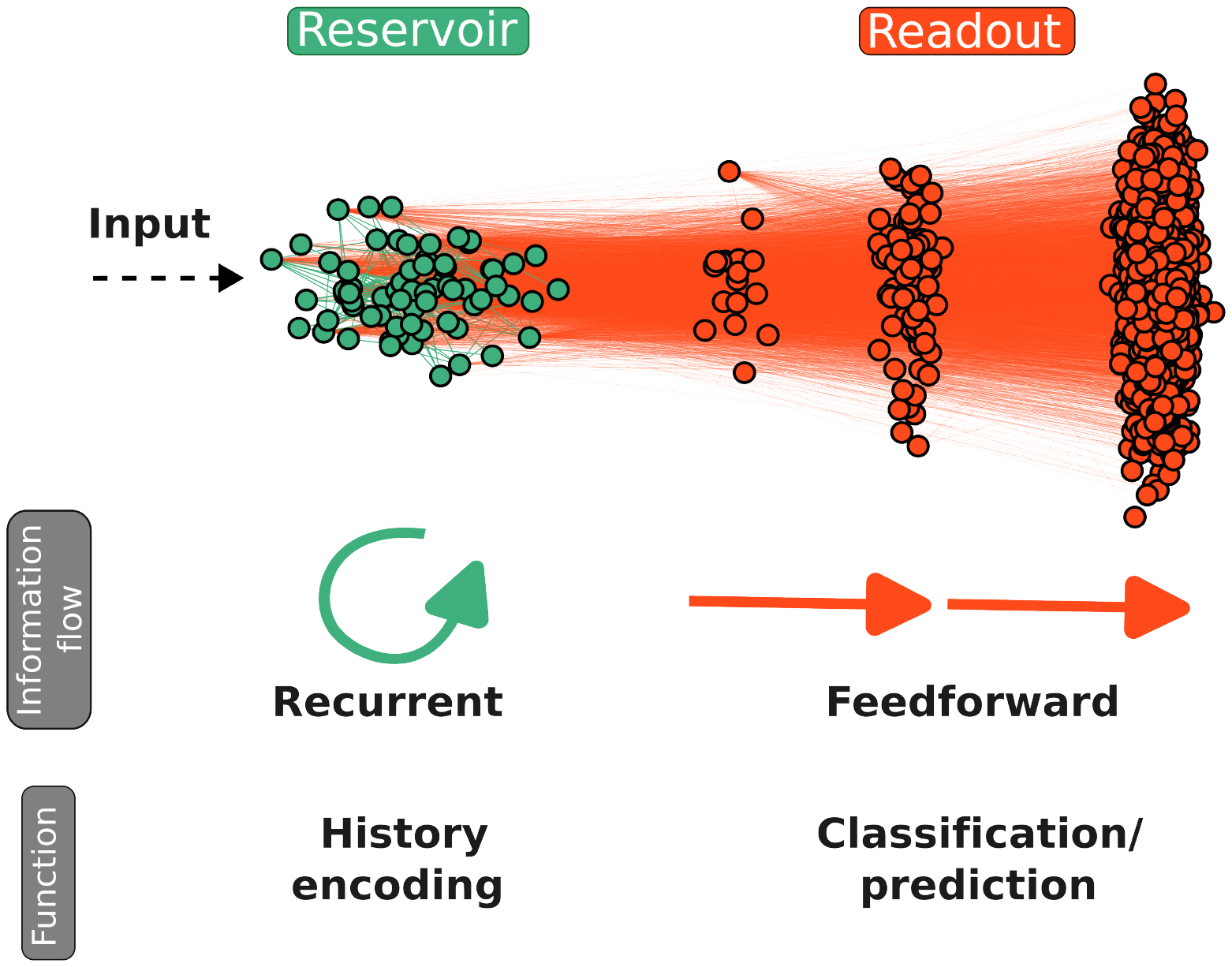}
}
\caption{\textbf{Structural and functional organization of the gene regulatory network of \textit{Escherichia coli}.}
The genes depicted in green constitute a recurrent subgraph featuring cyclic paths (corresponding to the reservoir) capable of storing the recent past within its dynamics. The nodes in red form a directed acyclic subgraph (corresponding to the readout) responsible for interpreting the state of the reservoir.
The network architecture shown here was obtained from the Ecocyc database \cite{Keseler2010}.
The representation is adapted from \citet{GabaldaSagarra2018}.}
\label{esquema_ecoli}  
\end{figure}

The main feature of RC is that memory encoding and prediction are handled by two separate network substructures, as depicted in Fig.~\ref{esquema_ecoli}.
First, a substructure with recurrent connections is needed to encode history: the reservoir. 
The reservoir serves two functions simultaneously:
\begin{enumerate}
    \item It performs a nonlinear expansion of the incoming signals (input) in a space of high dimensionality.
    \item It provides the system with memory by enabling it to retain information from its recent past.
\end{enumerate}
The second substructure shown in Fig.~\ref{esquema_ecoli} (red nodes) is the readout, located downstream of the reservoir and exhibiting a strictly feedforward architecture. This substructure combines the record of the immediate past encoded in the reservoir with the present input of the system, to generate predictions or perform classifications. This separation of roles allows to focus the training process (\textit{i.e.} the modification of the network's synaptic strengths so that the desired output is produced) solely on the readout \cite{Jaeger2004}.

In the classical RC approach, also known as an echo-state network (ESN), the reservoir is a random network with fixed (randomly set) weights, and solely the readout is trained.
Since feedforward architectures are much easier to train than recurrent ones, the RC framework combines the computational power of a recurrent network with the training simplicity of a feedforward network \cite{Buonomano2009}.
RC has demonstrated outstanding performance across various benchmark tasks \cite{Jaeger2007,Triefenbach2010PhonemeRW,Verstraeten2006ReservoirbasedTF}.

The simplicity of the network architecture underlying RC, and the evidence that natural gene regulatory networks exhibit this structure, support the expectation that recurrent biological networks might underlie the notable information-processing capabilities of living systems.
In the following section we introduce a basic mathematical description of how recurrent networks can compute.
Next we discuss the potential application of this paradigm in biological systems.
Finally we look into the evolutionary aspects of this natural computation framework. 

\section{How recurrent biological networks compute}

\subsection{Basic description}

In what follows, we focus on gene regulatory networks as information-processing substrates, although the same mathematical description can be applied to other biological networks (such as neuronal networks, which in fact are the ones most commonly considered in the literature within this context).
The dynamics of a gene regulatory network can be represented mathematically using models similar to those typically employed to describe artificial neural networks (ANNs).
Specifically, the expression level of a set of $N$ genes at a given time $t$ can be denoted by a vector $\mathbf{x}_t\in \mathbb{R}^N$, where $t=1\ldots T$ are a set of discrete time points at which the network activity is being monitored.
The expression of a gene depends on the levels of the genes that regulate it, and potentially on other inputs $\mathbf{u}_t\in \mathbb{R}^{N_\mathrm{in}}$ coming from outside the network.
This can be represented by the following update rule:
\begin{equation} 
    \mathbf{x}_{t}  = \mathbf{f}(W\mathbf{x}_{t-1} + W^{in}\mathbf{u}_t),
\label{res_dyn}
\end{equation}
where $\mathbf{f}$ is an element-wise activation function whose argument is a weigthed sum of the expression level of the regulating genes and the external inputs.
The effect of those regulating elements is represented by the weight matrices $W\in \mathbb{R}^{N\times N}$ and $W^{in}\in \mathbb{R}^{N\times N_{in}}$, respectively.
In the ANN literature, the activation function $\mathbf{f}$ typically takes for the form of a sigmoid (e.g. a hyperbolic tangent), which in fact is usually an excellent representation of the activation of a gene by its inputs \cite{carbonell2014bottom,rosenfeld2005gene}.

\begin{figure}[htbp]
\resizebox{0.5\textwidth}{!}{%
 \includegraphics{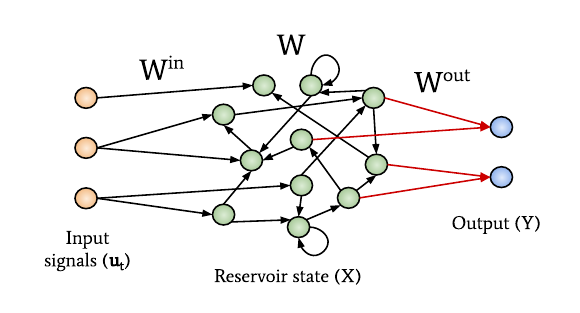}
}
\caption{\textbf{An echo state network}. The upstream genes (orange) send the external inputs ($\mathbf{u}_t$) to genes of the reservoir (green) with strengths given by the input weight matrix $W^{in}$.
The reservoir genes are coupled to each other by recurrent connections defined by the matrix $W$.
Finally, the readout genes (blue) read the information encoded in the reservoir to make decisions.
Only the readout weights $W^{out}$ (red arrows) are trained, and this in done in such a way that the output $Y$ approximates the target output signal(s).}
\label{esquema_res}  
\end{figure}

There is no constraint in the architecture of the network represented by the connectivity matrix $W$.
In particular, there is no need for the network to be feedforward, as is commonly the case in basic ANN (including deep-network) architectures.
In fact we consider here that the network underlying $W$ is recurrent, as in the ESN architectures used in RC \cite{Jaeger2004}.
The input signals $\mathbf{u}_t$ correspond to genes (or other biochemical elements) that are fully upstream of our network (orange nodes in Fig. \ref{esquema_res}).
In turn, the network output is given by a weighted sum of the states of the reservoir genes:
\begin{equation}
    \mathbf{y}_{t}=W^{out}\mathbf{x}_{t}
\end{equation}
where $\mathbf{y}_{t}\in \mathbb{R}^{N_{out}}$ is a vector of all predicted outputs, and $W^{out} \in \mathbb{R}^{N_{out}\times N}$ is the weight matrix for the reservoir-readout connections, as schematized in Fig. \ref{esquema_res}. 

The values of the readout weights $W^{out}$ determine the computational capabilities of the network.
In the artificial neural-network context those weights are trained in a supervised manner, to ensure that the network output $Y$ reproduces a given target output in the presence of the corresponding input.
This can be done very efficiently using a matrix operation such as ridge regression:
\begin{equation}
    W^{out} = Y^{\mathrm{target}}X^{T}(XX^{T} + \gamma ^{2}I)^{-1},
\label{ridge}
\end{equation}
where $Y^{\mathrm{target}}\in \mathbb{R}^{N_{out}\times T}$ is a matrix with all
expected/target outputs over time, $X \in \mathbb{R}^{N\times T}$ is a matrix with the state of the reservoir nodes over time, $X^{T}$ is the transpose of $X$, $I$ is the identity matrix, and $\gamma$ is a regularization coefficient, which helps avoid overfitting (see Sec. 8 of \citet{Lukoeviius2009} for details). 

It is worth noting that the ESN approach requires tuning the network appropriately.
Specifically, the recurrent subgraph should comply with the \textit{echo state property}: its state $\mathbf{x}_{t}$ needs to be exclusively determined by the fading memory of the incoming signal $\mathbf{u}_{t}$ \cite{Jaeger2007}.
To put it differently, the reservoir state $\mathbf{x}_{t}$ should remain independent of the initial conditions that were present before the input $\mathbf{u}_{t}$, if we consider a long enough $\mathbf{u}_{t}$ signal. Empirically, it has been seen that if the spectral radius (defined as the maximum absolute eigenvalue of the reservoir connectivity matrix $W$) is smaller than unity, then the \textit{echo state property} is satisfied.
For large enough spectral radii, reservoirs exhibit multiple fixed points or periodic, even chaotic attractors
(when the reservoir is sufficiently nonlinear).
This violates the echo state property and prevent efficient training.

\subsection{Alternative training algorithms}

The echo-state property described above implies that the reservoir is quiescent in the absence of external inputs.
However, biological networks usually display spontaneous activity characterized by intricate and irregular patterns. 
In fact, recurrent networks with chaotic self-sustained dynamics have been found to be highly effective for computational purposes \cite{Sussillo2009}.
In particular, chaotic networks exhibit faster training times and generate outputs that are more accurate and robust compared to non-chaotic networks.
Given the inability of the RC paradigm to operate in these conditions, a new training algorithm is needed.
For that purpose \citet{Sussillo2009} introduced a learning algorithm, called FORCE (First-Order Reduced and Controlled Error), that enables training the network even when it operates in a chaotic regime.
Interestingly, this algorithm is more biologically realistic than the one used in ESNs \cite{Jaeger2004}.
Given the occurrence of self-sustained dynamics in natural networks, and the relative realism of the FORCE algorithm for potentially training those networks, we discuss it in some detail in what follows.

\begin{figure}[htbp]
\resizebox{0.4\textwidth}{!}{%
 \includegraphics{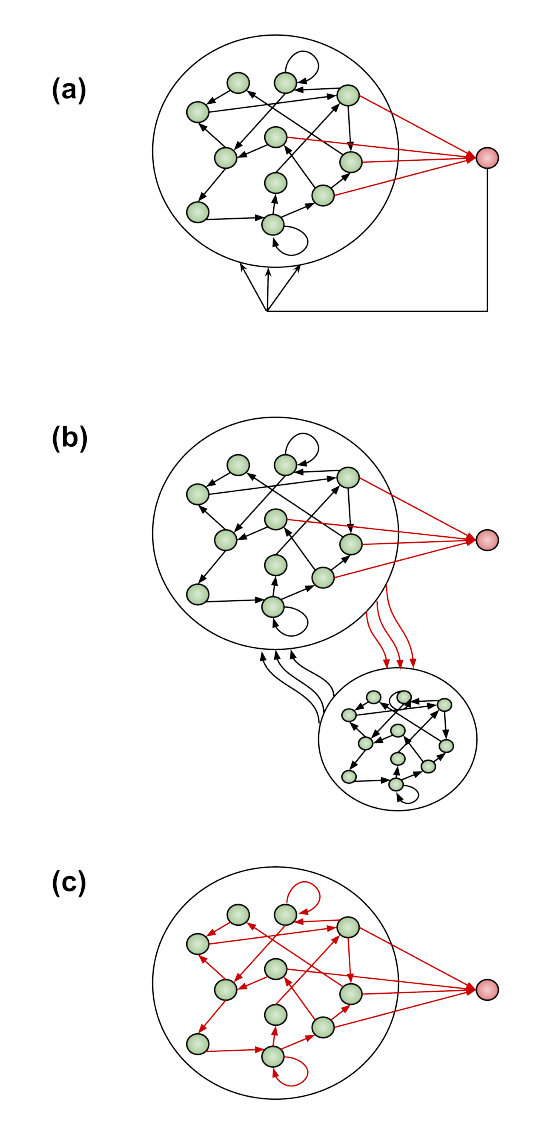}
}
\caption{\textbf{Network architectures for the FORCE algorithm \cite{Sussillo2009}}. In all cases, a reservoir (green nodes) drives a linear
readout unit (red node). Weights are only changed for the connections highlighted in red. (a) The readout feeds back into the reservoir. (b) A separate recurrent
network (smaller circled network) feeds back into the reservoir.
(c) No additional feedback (beyond the recurrence in the reservoir) is needed, and all connections are trained.
}
\label{esquema_sussillo}  
\end{figure}

To be operational, the learning algorithm must induce in the network a transition from chaotic to non-chaotic dynamics during training, and that is exactly what FORCE does. Specifically, \citet{Sussillo2009} show that the chaotic dynamics during training can be quenched by using either external feedback loops (Fig. \ref{esquema_sussillo}a) or internal synaptic modifications (Figs. \ref{esquema_sussillo}b and \ref{esquema_sussillo}c).
All three schemes lead to the generation of intricate yet controlled outputs. 
Driven by any of these two strategies, the key for suppressing chaos is making strong and rapid synaptic modifications.
In this regard, FORCE learning works in a manner distinct from traditional neural network training.
Typically, conventional training involves a series of adjustments aimed at gradually diminishing initially large errors in the network output.
The FORCE algorithm, in turn, reduces significantly the error's size at the initial weight update, and maintains the errors small during the rest of the training procedure. In order to achieve this, the method makes use of the recursive least-squares (RLS) algorithm \cite{haykin} as the update rule for the trained weights (see \citet{Sussillo2009} for details). Further modification becomes unnecessary upon completion of the training, and the network can autonomously produce the desired output.

Finally, it is worth emphasizing that the FORCE algorithm allows us to train not only the readout weights of the network, such as in the original RC framework (Fig. \ref{esquema_sussillo}a), but also RNNs with the architectures shown in Figs. \ref{esquema_sussillo}b and \ref{esquema_sussillo}c.
In these architectures, training affects not only the readout weights but also the internal synaptic connections. These networks resemble are more realistic biologically than that of Fig. \ref{esquema_sussillo}a, where feedback is only mediated by the readout unit.
By enabling training of different architectures, the FORCE algorithm extends the original RC framework, and gives more flexibility and biological feasibility to the concept of computation through recurrence. 

\subsection{Using dynamical systems theory to understand recurrence-based computations}

Since supervised training dictates the computation to be performed without constraining how the network executes this computation, the exact method through which trained RNNs achieve their intended functions remains an unanswered question.
In essence, the resulting networks have predominantly been regarded as black boxes. This differs from network models deliberately designed to implement specific established mechanisms, such as those proposed, for instance, by \citet{Hopfield1982} and \citet{Wang2008}.
One way of ``opening the black box'', proposed by \citet{Sussillo2013}, is to treat the network as a nonlinear dynamical system.

\citet{Sussillo2013} explored not only the existence of stable and unstable fixed
points and their linearized dynamics, but also the linearization in regions of phase space that are not genuine fixed points, but rather points where the system transitions very slowly (slow points).
They also present a straightforward optimization algorithm for finding such areas. 
Linearization around each one of the fixed and slow points provides a mechanistic description of how high-dimensional RNNs implement their computations \cite{Sussillo2013}. This final point is crucial for formulating hypotheses on how the task could potentially be solved by the biological network. Therein lies the importance of this dynamical-system based approach.

Optimization tools dependent on predefined analytic derivatives, such as those of the hyperbolic tangent function, are needed in order to find fixed and slow points of a RNN (see e.g. Section 3.2 of \citet{Sussillo2013} for details). Algorithms to find fixed points have been developed for simple (Elman) RNNs \cite{Bianchi2017}, and Hopfield-like RNN architectures \cite{Sussillo2013,katz_hop}, but can be inconvenient for complicated models, or for studies that
involve a combination of different RNN models. 
In this sense, a Tensorflow toolbox designed for determining fixed points and linearized dynamics in recurrent networks was introduced by \citet{Golub2018}. This software profits from TensorFlow's highly efficient automatic differentiation back-end to execute the necessary optimizations and compute Jacobians, essential for the identification and characterization of fixed points.

\section{Recurrent networks in biology}

\subsection{The mammalian cerebral cortex}

Cortical circuits were inspirational to \citet{Maassoriginal} when conceiving LSMs.
It has been proposed that cortical networks are likely to operate following an RC paradigm, or at least that they incorporate most RC principles.
One indication of this is the observable variation in both morphology and functionality among neurons across the cortex. Heterogeneity in neural characteristics has been observed in cortical neurons, the visual cortex and the retina, as well as in networks cultured \textit{in vitro} \cite{Haeusler2006,Bernacchia2011,Nikoli2009,Dranias2013,Ju2015,Marre2015}. In this direction, RC-like computation can make use of a very heterogeneous dynamics, that would arise from this diversity of neural components. In addition, cortical networks present a highly recursive nature \cite{Maass2016}.
It is unclear, however, to what extent recurrence plays a role in computation by cortical circuits, and efforts are under way to address this issue \cite{Haeusler2006,Bernacchia2011}. In what follows, we review current work in this direction.

\subsubsection*{Cortico-striatal system}
\citet{Domineyal1995} developed a RC-based model of the cortico-striatal system to shed light on the mechanisms underlying the behavior of the oculomotor system in different contexts \cite{Dominey1995}.
The model consists of two sub-networks, following the established organization of the oculomotor system of primates in cortical and subcortical structures. The first sub-network corresponds to the prefrontal cortex, which is responsible for a variety of cognitive tasks ranging from decision making to the regulation of social behavior \cite{fuster2015prefrontal}. The second sub-network corresponds to the striatum, which is a subcortical area. 

In the model of \citet{Domineyal1995}, the prefrontal sub-network has fixed recurrent connections between its neurons; while synapses from this structure to the striatum are modifiable \cite{Tanaka2019}.
The neural activity in the prefrontal network generates unique patterns of activity in response to the sequential visual inputs. These state-encoding patterns are then linked to the appropriate outputs (representing the oculomotor movements) by the striatum.

Later, \citet{Dominey2000} extended this model to the context of language processing in infants.
Specifically, the authors showed that the model described previously (which is also known as a temporal recurrent network) simulates the capabilities of babies to learn the sequential and rhythmic structure of language. For instance, the network can effectively encode sequential and temporal patterns, which can in turn be linked to specific behaviors in the presence of appropriate stimuli.
To reproduce the ability of babies to extract abstract structure from sequential inputs, the authors added a second recurrent network with short-term memory.
These works illustrate how inherent representational abilities of the structure of language might emerge from a shared neural architecture, separate from what is needed for representing abstract structures.

Augmenting the cortico-striatal model discussed above with a neurophysiologically grounded language model enables the resulting system to learn grammatical structures \cite{Dominey2009,Domineyetal2009}.
Furthermore, the efficacy of the learning algorithm can be increased by combining the model with an RC framework \cite{Hinaut2013}. The detailed history and the developments of the cortico-striatal model was reviewed by \citet{Dominey2013}.

\subsubsection*{Visual cortex} 

The visual system provides an excellent example of the need of processing time-dependent information.
\citet{Nikoli2009} obtained experimental evidence showing that the primary visual cortex of cats exhibits fading memory, which is one of the core features of ESNs.
Specifically, \textit{in vivo} spiking activities of 60–100 neurons were obtained for a sequence of different visual stimuli (letters).
The authors demonstrated that the neuronal responses that they recorded at a given time had information about the letter presented at that moment but also, interestingly, about the letters presented right before. In other words, the population activity was shown to encode both the present and the recent past.
As further verification, the authors used a network of linear integrate-and-fire  neurons as a readout, to perform a basic linear classification that enabled them to identify various visual stimuli reliably from the spike trains of the recorded neurons (representing reservoir states) \cite{Nikoli2009}.

The emerging viewpoint discussed above suggests that the visual cortex can process sequential inputs by engaging in time-dependent computations that provide the cell with the ability to store its recent past. This stands in stark contrast to the traditional belief that the brain process information locally in time, on a frame-by-frame basis, through feedforward network architectures \cite{Serre2005ATO} that are not well suited to process dynamic visual scenes. 

\subsection{Gene regulatory networks as reservoirs}

Biological computation through recurrence does not necessarily need to arise in neural substrates, but can be supported by other biological networks.
In this section, we review two example studies involving gene regulatory networks (GRNs).
The first study looks into the ability of \textit{Escherichia coli} to discriminate between time-dependent series of external chemical signals.
The second one studies the memory capacities of the GRNs of five evolutionary distant organisms.

\subsubsection*{Classification of chemical inputs through  \textit{E. coli}'s GRN}

\citet{bacteria2007} proposed that the gene regulatory network (GRN) of the bacterium \textit{Escherichia coli} can act as a computational reservoir, receiving input signals from a sequence of chemical stimuli and acting upon a set of target proteins as an output.
Specifically, the authors exposed a liquid culture of \textit{E. coli} cells to sequences of inputs that represent different combinations of temperature and chemical conditions of the environment.
Subsequently they studied the transcriptional state of the GRN using microarrays, to test whether the network was capable of making complex perceptual classification of the different inputs into desired classes. Numerical simulations of the aforementioned experimental setup showed that the RC framework could classify inputs with good performance, and perform logic operations such as XOR integration.

This work motivated an experimental study in which \citet{Didovyk2014} employed genetic engineering techniques to manipulate bacterial cell cultures, aiming to create a range of basic classifiers capable of distinguishing between various chemical inputs. The outputs could be further combined to solve more intricate classification problems.

\subsubsection*{Memory encoding through recurrence in GRNs}

The recurrence in the previous example was mostly caused by sparse self-regulation of a few genes, which was enough for the tasks proposed, which did not require memory.
Fully recurrent topologies are necessary, however, for processing time-dependent
information. 
In this sense, \citet{GabaldaSagarra2018} studied the GRNs of five different organisms (from bacteria to humans) in standard memory-demanding tasks.
Using public datasets, the authors identified that all five GRNs contained a recurrent reservoir formed by a small number of genes, while the majority of the genes in the networks were located downstream of the recurrent sub-graph, forming a feedforward network that can act as a readout, interpreting the high-dimensional dynamical state of the reservoir. In that way, the structure of GRNs are consistent with the echo-state network paradigm, providing the organism with a memory mechanism necessary to process temporal information.

Another important remark is that the training schemes used traditionally in the RC paradigm in the ANN context, such as ridge regression (Eq. \ref{ridge}), are not biologically realistic.
To address this issue \citet{GabaldaSagarra2018} studied whether an evolutionary process could successfully train the readouts.
Their results showed that the output weights could indeed be tuned through an evolutionary algorithm, mimicking the way in which evolution would act upon the GRNs to enable computation.

\section{Evolutionary aspects of recurrence-based computations}


Besides the role of evolutionary processes in the feasible training of the output readouts in the RC framework, there are other relevant evolutionary aspects of RC that are worth highlighting here.
In particular, adequate design properties are needed to produce operating reservoirs, as discussed by \citet{Seoane2019}.
As for all biological strategies, it is expected that critical design decisions are influenced by evolutionary constraints.
In the following we review the evolutionary constraints that underlie optimal operating reservoirs. 

When RC was initially introduced, it was shown mathematically that the reservoir needs to have two fundamental attributes for the proper implementation of the paradigm \cite{Maassoriginal}. Firstly, the dynamics of the reservoir needs to be sufficiently high-dimensional to effectively \textit{separate} training inputs across a variety of tasks. Secondly, the system should be able to \textit{generalize} its classification abilities to new examples.
This would entail mapping similar inputs to nearby regions within the dynamical reservoir's phase space.

\citet{Legenstein2007} analyzed how reservoir dynamics contributes to those two requirements, showing that separability and generalization are anticorrelated with each other.
Recurrent networks with simple attractor dynamics, for instance, have substantial generalization capabilities (because even under noise, many inputs give rise to similar dynamics in the reservoir's phase space), they exhibit low separability.
When recurrent networks operate in the chaotic regime, in contrast, the dynamics becomes highly sensitive to initial conditions, which strongly impairs generalization, while separability remains large due to the nonlinear and high-dimensional character of the system. 

Consequently, to fulfill simultaneously the conditions of separability and generalization, recurrent networks need to operate near the so-called \textit{edge of chaos}, at the boundary between order and chaos.
This critical region represents a compromise between attractor dynamics, in which perturbations (even relatively large ones) are washed away as the network approaches equilibrium, providing generalization, and chaotic dynamics, where perturbations (even very small ones) are amplified, providing separability.
This requirement is consistent with the large amount of evidence reported during the past two decades showing that the brain operates near criticality\cite{beggs_criticality_2007,beggs_neuronal_2003,r_chialvo_critical_2004}. 

Following this line of reasoning, criticality has been used in recent years as a global constraint in the design of optimal reservoirs, employing quantifiers such as Lyapunov exponents \cite{Legenstein2007,Bertschinger2004,Toyoizumi2011,Boedecker2011,Bianchi2018} and Fisher information \cite{Livi2018} of reservoir dynamics. 
Taken together, these studies highlight the advantages of criticality for information processing, and suggest powerful constraints that could have played a relevant role in evolution of the computational capabilities of recurrent biological networks. 

\section{Final remarks}

Living organisms are faced with the ongoing challenge of processing multiple simultaneous signals coming from their dynamical environment.
In spite of the ubiquity of this requirement, a general conceptual framework of how biological systems process temporal information is still lacking.
The intrinsically dynamical nature of recurrent networks allows in a natural way the integration of the present conditions faced by the living organism with its (recent) past, making the concept of recurrence-based computation an appealing initial step towards a conceptual model of temporal information processing in living beings.

In spite of their conceptual appeal, validating experimentally the role of recurrence-based computations in biological systems is still challenging.
For example, while the existence of recurrent architectures in cortical brain regions has been thoroughly studied, as described above, research linking systematically the dynamics of recurrent cortical networks with time-dependent tasks is still needed. 
This will require a comprehensive analysis of cognitive systems at the structural, dynamical, and functional levels. 

Model animals could be a suitable starting point for that goal.
For instance, the fully mapped connectome of the roundworm \textit{C. elegans} has been shown to be consistent with a network architecture that supports recurrence-based computations.
When the dynamics of this network is compared with experimental observations, the system is seen to function in an input-driven chaotic regime that supports long-term memory via the phenomenon of generalized chaos synchronization \cite{Casal2020}.
Future interdisciplinary research should bridge the existing substantial literature on dynamical systems with our emerging knowledge of human and animal brains, together with our increasing understanding of the computational properties of recurrent neural networks.
Given the fact that adaptability to changing environments is one of the main defining characteristics of living systems, the synergistic interaction between applied mathematics, neuroscience and machine learning discussed in this review can be expected to provide us with a general principle of biological information processing that has so far been elusive.

\section*{Acknowledgments}

This work was supported by project PID2021-127311NB-I00 / MICIN / AEI / 10.13039 / 501100 011033 / FEDER-UE, financed by the Spanish Ministry of Science and Innovation, the Spanish State Research Agency and the European Regional Development Fund (FEDER).
Financial support was also provided by the Maria de Maeztu Programme for Units of Excellence in R\&D (Spanish State Research Agency, project CEX2018-000792-M), and by the ICREA Academia programme.
M.S.V. is supported by a PhD fellowship from the Ag\`encia de Gesti\'o d'Ajuts Universitaris i de Recerca (AGAUR) from the Generalitat de Catalunya (grant 2021-FI-B-00408).

%

\end{document}